# Are EU's Climate and Energy Package 20-20-20 targets achievable and compatible? Evidence from the impact of renewables on electricity prices

Juan Ignacio Peña[a], and Rosa Rodríguez[b]

## Abstract

This paper studies the realizability and compatibility of the three CEP2020 targets, focusing on electricity prices. We study the impact of renewables and other fundamental determinants on wholesale and household retail electricity prices in ten EU countries from 2008 to 2016. Increases in production from renewables decrease wholesale electricity prices in all countries. As decreases in prices should promote consumption, an apparent contradiction emerges between the target of an increase in renewables and the target of a reduction in consumption. However, the impact of renewables on the non-energy part of household wholesale electricity prices is positive in six countries. Therefore, decreases in wholesale prices, that may compromise the CEP2020 target of decrease in consumption, do not necessarily translate into lower household retail prices. Monte Carlo simulations suggest that the probability of achieving CEP's target of reductions in GHG emissions for 2020 is lower than 1% in Austria, Portugal, and Spain. In horizon 2030, Austria, France, Germany, Portugal, and Spain have probabilities lower than 1% of achieving the GHG emissions target. Finland and France present success probabilities lower than 1% on the national targets of renewable sources for 2020 and 2030 as do Austria and Spain with reductions in electricity consumption.

*a*. Corresponding author. Universidad Carlos III de Madrid, Department of Business Administration, c/ Madrid 126, 28903 Getafe (Madrid, Spain). ypenya@eco.uc3m.es; *b.* Universidad Carlos III de Madrid, Department of Business Administration, c/ Madrid 126, 28903 Getafe (Madrid, Spain). rosa.rodriguez@uc3m.es.

# 1. Introduction

The 20-20-20 Climate and Energy Package (CEP2020 henceforth) is binding legislation adopted in late 2008[1], setting three targets. The first target requiring EU's Member States to cut greenhouse gas emissions (GHG) by 20% relative to 1990 levels, for the year 2020. The second target sets an EU-wide share of 20% of gross final energy consumption from Renewable Energy Sources (RES), although mandatory national targets vary from 10% in Malta to 49% in Sweden. CEP2020 includes an additional non-binding third target to cut gross primary energy consumption by 20% by 2020 compared to projections published in 2007 and in 2009 (EC, 2007 and 2009). In October 2014, Member States set targets for 2030: cut greenhouse gas emissions by 40%, produce 27% of energy from RES and cut energy consumption by 27%.

This paper studies the realizability and compatibility of the three CEP2020 targets, focusing on electricity markets because electricity generation is the single largest emitter of carbon to the atmosphere. Our first contribution to the literature is to study the role of renewables generation as a determinant of wholesale prices in ten EU countries consuming over 95% of electricity in the EU, in the period 2008-2016. This merit-order effect has been studied in the literature. Würzburg, et al. (2013, Table 1) present results of seventeen studies in seven different countries[2], thirteen of them reporting a negative impact of renewables generation on wholesale spot prices[3]. The decrease in wholesale electricity prices for each GW of additional renewable power ranges from 2 €/MWh in Spain (Gelabert, et al., 2011) to 9.9 €/MWh in Ireland (O'Mahoney and Denny, 2013), but in Denmark, Jonsson et al. (2010) document an 40% electricity price change between low-wind and high-wind circumstances.

---

[1] https://ec.europa.eu/clima/policies/strategies_en
[2] Ten studies are based on simulations and seven on empirical analysis. The countries are Germany, Denmark, Spain, UK, Norpool, Australia, and the USA.
[3] Three studies are inconclusive, and one study presents a long-term positive impact.

Our contribution to this strand of literature is twofold. First, while earlier studies focused in one or two markets and use different methodologies, methods, data and objectives making hard to draw general conclusions, we compare ten different markets using a common method and data sample, so allowing for a more comprehensive assessment of the merit-order effect. Second, by using monthly data, we complement the evidence of many other studies, usually based on daily data, Gelabert, et al. (2011).

Our second contribution is to analyze the impact of renewables generation of household retail prices in the EU. Moreno et al. (2012) and Pereira da Silva and Cerqueira (2017) report a positive effect of renewables generation on the energy component of household retail prices (apart from taxes and levies). Household retail prices contain two components: the cost of energy (including transmission costs) and the government wedge (taxes and public policy cost). Increasing generation from renewables should have a negative impact on the former and a positive impact on the latter. However, in the two studies mentioned above the dependent variable is the household retail price *excluding* taxes and levies. These prices should be largely determined by the cost of energy and the impact of renewables should be negative. As a way of dealing with this conundrum, we posit another approach based on subtracting wholesale prices from household prices, getting an estimate of the "government wedge" part. This procedure allows us to analyze the impact of renewables on the non-energy part and the impact of the balance of the national budget as a proxy for the financing needs of the government.

Our third contribution is to carry out simulations for the 2020 and 2030 horizons to study to what extent the three targets set out in 2009 and 2014 are compatible and likely to be achieved. In doing so, we extend Liobikienė and Butkus (2017) by including a wider array of models, and a more comprehensive discussion of scenarios.

The empirical analysis suggests that the impact of renewables on wholesale prices is negative and relevant in all countries, but the size of the impact varies significantly between countries. For instance, an increase of one percentage point in the proportion of renewables over gross generation decreases wholesale prices by 0.167 €/MWh in Austria but by 0.506 €/MWh in Spain. Consumption, Fuel prices (oil, coal, gas), carbon prices, temperature and net balance between exports and imports are relevant variables in some countries but not in others. Imposing common responses to the impact of the fundamental variables on electricity prices, as do some panel data models, EC (2015), may produce misleading interpretations and conclusions. The impact of renewables on the non-energy part of household wholesale electricity prices is positive and significant in six countries, and the Government deficit-to-GDP ratio lagged one period is negative and significant in three countries.

Simulation-based forecasts suggest that the probability of achieving EU's CEP target of reductions in greenhouse gas emissions and electricity consumption for horizons 2020 and 2030 is lower than 1% in Austria, Portugal, and Spain. In horizon 2030, Austria, France, Germany, Portugal and Spain present probabilities lower than 1% of achieving the cut in GHG emissions consistent with the CEP target. Finland and France present probabilities lower than 1% regarding the national targets of renewable sources for horizons 2020 and 2030 as do Austria and Spain with reductions in electricity consumption.

The rest of this paper is organized as follows. Section 2 reviews the literature. After describing the methodology in Section 3, we present data in Section 4. Section 5 discusses empirical results. Section 6 concludes.

## 2. Literature Review

Extant literature has discussed the question of inconsistencies in, and achievability of, the current CEP2020 targets. Böhringer, et al. (2009a) highlight the potential inconsistency of using three instruments and one target, in conflict with the classical suggestion in Tinbergen (1952). On the other hand, Del Rio (2017) supports the combination of an ETS and renewable energy targets as intermediate steps toward the key target of reductions in GHG emissions because of different policy goals and market failures. Böhringer, et al. (2009b) suggest the inefficiencies in EU's climate policy lead to a cost a 100–125% too high[4]. Tol (2012) gauges the benefit–cost ratio of the EU emissions targets for 2020 to be 1/30 and Flues at al. (2014) suggest that excess costs can reach up to 1.2 Billion Euro annually. Böhringer et al. (2016) argue that the multiple instruments used in the EU to curb GHG emissions generate an excess cost. Unintended consequences of the CEP2020 package are analyzed in Böhringer and Rosendahl (2010) who show adding a minimum renewable share target to an existing ETS promotes production by the dirtiest power generation technology. A related point is made by Sinn (2015) pointing out that policies like CEP2020 aimed at limiting conventional fuels (e.g. oil, coal) induce fuel owners to bring their sales forward to the present, depressing current market prices and increasing demand, boosting GHG emissions and giving rise to the "Green paradox". Although Knopf et al. (2015) remark the absence of a detailed analysis of the consistency between the individual targets of CEP2020, Helm (2012) points out that the CEP2020 renewables target supports a few intermittent and expensive technologies (wind, solar). As they include more renewables in the system, carbon prices should fall. As CEP2020 fixes the total quantity of emissions and then imposes a renewables target, the effect on GHG emissions of both (contradictory) directives are likely to be minimal. Helm (2014) stresses the unsustainability of CEP2020 because it increases energy costs, make the EU less competitive, does not guarantee secure supply, and concentrates on taxing carbon production instead of carbon consumption. Based on these concerns the EU may need a thorough reassessment of its climate and energy policies.

---

[4] This estimate is very robust, see Hermeling, et al. (2013).

Liobikienė and Butkus (2017) estimate that almost all EU countries will achieve their 2020 GHG targets. However, several countries may not achieve the targets of energy consumption nor the targets of RES generation. Khanam et al. (2017) surveyed experts from EU countries, and report 49% saying the renewable energy target will not be met, 60% do not trust the GHG cut target will be fulfilled, and 85% saying the EU's energy efficiency target will not be met.

## 3. Methodology

We estimate the following (1) panel multivariate linear regression model to study the impact of renewables generation and other fundamental variables on electricity wholesale prices.

$$p_{t,j} = \beta_{0,j} RES_{t,j} + \sum_{i=1}^{K} \beta_{j,i} X_{t,j,i} + \eta_{t,j} \quad (1)$$

Where the dependent variable $p_{j,t}$ is the wholesale electricity price in country $j$ at time $t$, $RES_{t,j}$ is the proportion of generation by renewables divided by gross generation and $X_{t,j,i}$ are $i=1,\ldots,K$ explanatory variables in $j=1,\ldots,J$ countries either stochastic (e.g. demand) or deterministic (e.g. fixed effects or dummy seasonal variables)[5]. An important point is whether the slope corresponding to an explanatory variable is the same for all countries. We check this assumption through the Wald test, where the null hypothesis is $\beta_{k,i} = \beta_{m,i} \, \forall \, k,m \in \{J\}$. The noise process $\eta_{t,j}$ follows an ARMA process

$$\phi(B)\eta_{t,j} = \theta(B)\sigma_{t,j}\varepsilon_{t,j} \qquad (2)$$

$$\phi(B) = 1 - \sum_{l=1}^{\phi}\phi_l B^l; \quad \theta(B) = 1 - \sum_{l=1}^{\theta}\theta_l B^l \qquad (3)$$

---
[5] We include two dummy variables, CWE corresponding to the Market Coupling in Central Western Europe in 11/2010 and NWE corresponding to the Market Coupling of Northwestern Europe in 2/2014 CWE and NEW are equal to zero before the corresponding date and equal to one afterwards.

Functions $\phi(B)$ and $\theta(B)$ are polynomial functions of the backshift operator B, $\sigma_{t,j}$ is the volatility and the innovation $\varepsilon_{t,j}$ is N(0, 1). The estimation method of (1)-(3) is Cross-section SUR, meaning a feasible GLS specification correcting for heteroskedasticity and contemporaneous correlation and the covariance matrix is estimated by means of Cross-section SUR Panel Corrected Standard Error (PCSE).

Explanatory variables $X_{t,j,i}$ are fundamental factors, Weron (2014): (i) Fuel prices, (ii) Weather variables (iii) Demand (consumption), (iv) Carbon prices, and (v) Difference between electricity exports and imports (net balance). The longer the time interval the more important fundamental factors become[6]. Table A1 in the Appendix summarizes the explanatory variables $X_{t,j,i}$.

Household retail prices contain two elements: the cost of energy (wholesale prices and network costs) and the "government wedge" (taxes and public policy costs). Moreno et al. (2012) and Pereira da Silva and Cerqueira (2017) explain EU household retail electricity prices (apart from taxes and levies) using several factors, including amount used, fuel prices, and the part of gross generation by renewables. Using a panel regression model and restricting the slope coefficient of all explanatory variables to be the same for every country, they report a positive effect of renewables generation on household retail prices. These results are intriguing because the impact of renewables on household prices should have a twofold effect[7]. On the one hand, renewables generation decreases the cost of the energy part because of the merit order effect. Therefore, we should see a negative coefficient in the corresponding regression. But renewable generation is financed by public policies through taxes and levies and so we should expect a positive relationship between renewables generation and the "government wedge" part. The ultimate impact of renewables on household prices will depend on the relative size of the two components of retail prices, a situation that is country-specific and varies over time[8]. However, in the two studies mentioned above the dependent variable is the household retail price apart from taxes and levies. Consequently, these prices should be largely determined by the cost of energy and the impact of renewables should be negative instead of the positive coefficients reported by both papers. In fact, we document a negative impact of renewables on wholesale electricity prices in Table 1, in agreement with many published studies.

---

[6] For instance, studies show fuel costs may not influence high-frequency electricity prices (Guirguis and Felder, 2004). In this paper we focus on a medium to long-term perspective. Monthly data provides a long sample size and smooths out the high volatility and spikes intra-day or daily data presents.

To shed light on this issue, a cleaner estimate of the impact of renewables on the non-energy component of household prices may be gauged by subtracting wholesale prices from household prices, getting an estimate of the "government wedge" part. This procedure allows us to analyze the impact of renewables on the non-energy component. In doing so, we need not include in the regression model variables associated specifically with the energy part (e.g. fuel prices, weather). But we should include variables correlated with the size of the government wedge. A natural explanatory variable is the balance of the national budget as a proxy for the financing needs of the government, and we include this in the following regression model. The regression is

$$(h_{t,j} - p_{t,j}) = \alpha_j + \delta_j RES_{t,j} + \varphi_j \nabla PB_{t-1,j} + \eta_{t,j} \quad (4)$$

---

[7] We should like to thank an anonymous referee for pointing this out.
[8] Robinson (2015) reports that in the second half of 2014, the government wedge component accounted for 29%, 33%, 38% and 45% of household retail prices in France, Germany, Italy and Spain, respectively. In the EU during the period 2008-2012, the energy component (wholesale prices) decreased by 4% and the government wedge component increased by 31%.

Where the dependent variable $(h_{j,t} - p_{t,j})$ is the household retail electricity price minus the wholesale price in country $j$ at time $t$, and the noise process $\eta_{t,j}$ follows (2)-(3). RES is the generation by renewables divided by the gross generation and $\nabla PB$ is the lagged change in government deficit-to-GDP ratio (positive if the government has a budget surplus). We expect a positive coefficient in the RES variable, as discussed above, and a negative coefficient in the $\nabla PB$ variable. The economic intuition for this is that a deteriorating fiscal position (negative change) incentives government to step-up its collecting activities, which may hint at increases in the non-energy part of household prices.

Third, we do Monte Carlo simulations for the evolution of greenhouse gas emissions (GHG), generation by renewables (RES) and electricity consumption (CONSM) in each country $j$ by 2020 and 2030. These simulation-based forecasts depend on univariate[9] models,

$$(1 - \rho_{1,j}B - \rho_{2,j}B^2)Y_{t,j} = \mu_j + \sigma_j \varepsilon_{t,j} \quad (5)$$
$$Y_{t,j} = \beta_{0,j} + \beta_{1,j}t + \sigma_j \varepsilon_{t,j} \quad (6)$$

where $Y_{t,j}$ ={GHG, RES, CONSM}, and $\varepsilon_{t,j}$ is a Niid(0,1) variable[10]. The reason for using these models is to describe the behavior of the data and, in particular, to find out if GHG emissions, electricity consumption, and share of RES exhibit an increasing or decreasing trend different from random behavior. In the case of the target of reductions in GHG emissions, we also fit a model in which the dependent variable is the first difference of the logarithm of GHG emissions and the explanatory variables are first differences of the logarithm of consumption and RES. The reason for doing this is to highlight the interactions between the rate of change of the three CEP2020 targets.

---

[9] Univariate models are (i) deterministic linear trend and (ii) AR (2). The first model has been used in the literature (Liobikienė and Butkus, 2017). As an alternative, the AR(2) model has been used extensively in the literature and allows for stochastic trends and cycles, Weron (2006).
[10] The simulation is done using yearly data, based on averages of daily data, and therefore the Normality assumption we use follows from the central limit theorem.

$$Log(\nabla GHG_{j,t}) = \beta_{j,0} + \beta_{j,1} Log(\nabla CONSM_{j,t}) + \beta_{j,2} Log(\nabla RES_{j,t}) + \varepsilon_{j,t} \quad (7)$$

This procedure allows as to do scenario analysis, considering the impact of different trajectories for the explanatory variables on GHG's growth rate. Additionally, we can check whether considering these interactions and scenarios produce different results in comparison with the results we got from simple models (5) and (6). The estimation method of (7) is the same as in (1). Based on the empirical distribution of the simulated paths, we compute the possibilities of compliance of the 2020 and 2030 targets for each country.

Current literature presents other competing models for the analysis of the impact of renewables on electricity prices and for the generation of scenarios, for instance, those based on a detailed representation of the electricity sector. Sensfuß et al. (2008) use a model of the German electricity market to run several simulations for situations with and without renewable production to find out the effect on wholesale prices. Weigt (2009) presents a market model, designed to minimize costs (including unit commitment and start-up costs), to estimate the differences in production costs and market prices caused by wind penetration. In contrast, in this paper, we focus on ex-post data on electricity prices and renewable capacity in several countries using a common econometric approach to compute the actual price effect of expanding renewables and simulating the trajectories of the three CEP2020 targets over time.

## 4. Data

We compute monthly electricity wholesale prices as averages of daily prices (one-day-ahead, baseload, €/MWh) from ten EU countries: Austria, Denmark, Finland, France, Germany, Greece, Italy, Portugal, Spain, and Sweden, amounting for 95% of total electricity consumption in the EU. Monthly data for explanatory variables are computed in the same way. Table A1 in the Appendix has data definitions and sources. The sample size is 1,080 data points from January 2008 to December 2016 and data providers are the corresponding electricity exchanges (pools). Preliminary data analysis shows that the wholesale price series display the typical electricity price features (Escribano et al., 2011). Panel A in Table A2 in the Appendix provides information on basic statistics for monthly price series. Panels B and C in Table A2 in the Appendix present summary statistic for wholesale and household retail biannual prices. Figures 1 and 2 show wholesale and retail prices.

[INSERT FIGURE 1 HERE]

[INSERT FIGURE 2 HERE]

The sample size is 180 biannual observations because the data provider (EUROSTAT) publishes these data in that frequency only. Household retail prices are on average about four times higher than wholesale prices, extreme cases being France (2.42) and Denmark (7.36).

## 5. Results and discussion

Several studies (e.g. Würzburg, et al., 2013) document that increasing generation from renewables is associated with decreases in wholesale prices, the well-known merit-order effect (Jensen and Skytte, 2002), so we expect a negative regression coefficient in the variable RES. We expect increases in consumption and fuel prices to be associated with increases in electricity wholesale prices. Increases in the carbon price ($CO_2$ emission allowances) should correlate with increases in prices. In countries with low average temperatures, the expected sign of the temperature variable is negative. Regarding net balance (*Netbal*), the difference between exports and imports of electricity, there are alternative situations. If domestic prices are below (above) prices in connected neighbor countries, is profitable to export (import) power to (from) other countries. Therefore, the impact of Netbal may vary. If in the domestic market there is high (low) demand and net balance is positive, we should see a positive (negative) correlation between domestic wholesale prices and Netbal. But, if in the domestic market there is high (low) demand and net balance is negative, we expect a negative (positive) correlation between wholesale prices and Netbal because net (exports) imports (increase) decrease domestic prices. We expect the impact of the dummy variables NWE and CWE, associated with market couplings, to be negative because market integration should decrease wholesale prices. Table 1 shows the estimation results of the model (1)- (3) using monthly data in the period from 1/2008 to 12/2016.

[INSERT TABLE 1 HERE]

The impact of RES on wholesale prices is negative and significant, as expected. However, the size of the impact varies between countries and the Wald test rejects the null hypothesis of an equal coefficient for all countries. For instance, an increase of one percentage point in the proportion of renewables over gross generation decreases wholesale prices by 0.167 €/MWh in Austria but by 0.506 €/MWh in Spain. Consumption presents a relevant effect in three cases with a positive sign, as expected. Fuel prices (oil, coal, gas) are significant in twelve cases. Increases in carbon prices correlate positively with increases in prices in two cases. Temperature has a negative and significant impact in four cases (Denmark, Finland, Sweden, and Germany) corresponding to countries with low average temperatures, as expected. Netbalance presents positive impact in three cases and negative impact in three cases. The dummy variables NWE and CWE are not significant, suggesting a lack of impact of market couplings on prices. Wald test suggests that the hypothesis of equality of coefficients between countries has a probability lower than 1%, excepting for carbon prices, with a probability of 1.8%. The VIF multicollinearity indicator is below 10 in all cases, suggesting a lack of harmful collinearity. Durbin-Watson statistic does not signal residual autocorrelation problems. The Hausman test statistic is 80.27 and the related p-value (approximately 0) lead to a rejection of the null hypothesis of strict exogeneity. Therefore, we conclude that the fixed effects model is more suitable than the random effects option.

The regression results of household retail prices (4) are shown in Table 2. the dependent variable $(h_{j,t} - p_{t,j})$ is the household retail electricity price minus the wholesale price in country $j$ at time $t$, and the noise process $\eta_{t,j}$ follows (2)-(3). Biannual data spans the period 2008-2016.

[INSERT TABLE 2 HERE]

The impact of renewables is positive as expected (except for Austria) and parameters are statistically relevant in six out of ten countries. In the same way to the case of wholesale prices, the size of the impact varies between countries and the Wald test rejects the null hypothesis of an equal coefficient for all countries. For instance, an increase of one percentage point in the part of renewables over gross generation increases the non-energy part of household prices by 0.22 €/MWh in Spain and by 1.99 €/MWh in Italy.

Government deficit-to-GDP ratio lagged one period is a salient explanatory variable in Finland, Portugal, and Spain, with the expected negative sign. This is direct evidence of the role that the fiscal position in a country may have on the size of the non-energy part of electricity retail prices[11]. The VIF multicollinearity indicator and Durbin-Watson statistic do not signal multicollinearity nor residual autocorrelation problems, respectively. The Hausman test statistic is 70.26 with a p-value roughly zero. So, data rejects the null hypothesis of strict exogeneity. Therefore, we conclude that the fixed effects model is more suitable than the random effects choice.

Next, we study the simulated distribution in greenhouse gas emissions (GHG), generation by renewables as proportion over gross generation (RES) and electricity consumption (CONSM) by 2020 and 2030.[12]

The first target in CEP2020 is a cut in GHG emissions of 20% in 2020 and 40% in 2030 compared to 1990. We use the deterministic trend model as in Liobikiene and Butkus (2017) and an AR(2)[13] model. The deterministic trend model fits data better and we talk about its results here[14]. Table 3 compares the simulated terminal values based on 100,000 trajectories

---

[11] Notice that in Table 1 we estimate 103 parameters with a sample size of 1,080 data points, so we have 10 subjects per variable (SPV). In Table 2 we estimate 21 parameters with 180 data points, so the SPV is 8. These SPV are well above of the amounts required to get accurately estimated regression coefficients, standard errors, confidence intervals, and R-squares (Austin and Steyerberg, 2015).

[12] Greenhouse gases data have been obtained from the European Environment Agency (EEA). To keep uniform the data sample used in the paper GHG, figures for the year 2016 have been computed. We have used the data of emissions in Carbone Dioxide for all countries, obtained from Eurostat, and we kept the proportion of $CO_2$ and non-$CO_2$ existing in the previous year.

[13] In Appendix A, Table A3 presents the results of the estimation and the test for residual correlation and heteroscedasticity.

[14] As a robustness check, we computed the probabilities of compliance for the 2020 and 2030 targets using the

against target values. The first two rows of Table 3 present the GHG emissions for each country in 1990 and the target value for 2020, got by decreasing by 20% the 1990 figure. The rows Q(1%,2020) and Q(50%,2020) contain the 1% and 50% quintile of the distribution of forecasts for horizon 2020 respectively, meaning that 99% and 50% of the simulated trajectories yield terminal values higher than this figure. For instance, with Austria, the 2020 target is 63,760 and the Q(1%,2020) is 70,739. Therefore, the model suggests that the probability of observing reductions in GHG emissions consistent with the CEP2020 target is lower than 1%. Portugal and Spain are in a similar situation. With 2030, Austria, France, Germany, Portugal and Spain present probabilities lower than 1% of achieving the cut in GHG emissions consistent with the CEP2020 target.

[INSERT TABLE 3 HERE]

To gain further insight into the probabilities of achieving the key target of reductions in GHG emissions, we do a Monte Carlo simulation based on model (7)[15]. To do this we need assumptions on the evolution of the explanatory variables in (7), RES and consumption. Table 4 has average growth rates of the three variables included in the model. The average rate of growth in GHG emissions ranges from -0.99% in Germany to -4.59% in Greece and is -2.2% for all countries together. Growth in the amount used ranges from -1.54% in Sweden to 0.90% in Austria and the average is -0.5% considering all countries. Average yearly increases in RES range from 1% in Sweden to 15% in Greece and the overall average is 6.7%.

.

---

AR(2) model. Results are in Table A5 in Appendix A. Results do not change in comparison with the ones got using the deterministic trend model. The 2020 target would not be achieved in the same three countries: Austria, Portugal and Spain. Regarding 2030, 7 out 10 countries will not achieve the cut in GHG emissions target.

[15] Results of estimating model (7) are in the Appendix.

[INSERT TABLE 4 HERE]

The results of the simulations based on model (7) are in Table 5.

[INSERT TABLE 5 HERE]

As an illustration, we consider three scenarios whose results are in panels A, B, and C of Table 5. In Panel A, the "business-as-usual" scenario, consumption and RES in each country increase at the average of the growth rate observed in the period 2008-2016. Considering all countries, the figures are -0.5% and 6.7% respectively. In Panel B, a "cut-in-consumption-bullish-but-RES-bearish" scenario, consumption decreases by 2% per year and RES increases only by 2% yearly. Finally, in Panel C, a "cut-in-consumption-but-RES-average" scenario, consumption decreases by 2% per year and RES increases by 6% yearly[16]. In Panel A of Table 5, we may see that Austria, Portugal, and Spain have less than 1% chance of achieving the 2020 and 2030 targets. Panels B and C suggest that those three countries will struggle to meet the 2020 target but may achieve the 2030 target. In summary, results from the three models (5), (6) and (7) concur that the probabilities of meeting 2020 (and perhaps 2030) GHG target are less than 1% for Austria, Portugal, and Spain.

The second target set by CEP2020 is 20% production by RES by 2020 and 27% by 2030. We model yearly production by renewables divided by gross generation by using models (5) and (6)[17]. The trend model fits data better, and we generate 100,000 trajectories and compute the 99% quantile of the empirical distribution of terminal values for horizons 2020 and 2030.

---

[16] We considered other alternative "pessimistic" scenarios in which consumption increases and RES stagnates. The results are available on request.
[17] Estimation results are not included to save space and are available in Appendix A in tables A6.

Results are in Table 6.[18]

[INSERT TABLE 6 HERE]

Finland and France present a probability lower than 1% of achieving the RES target in 2020 or 2030. As before, results assume of no significant policy changes. Perhaps, when modelling the growth pattern of RES, the logistic model may be an alternative to the linear method used in this paper. Hansen et al. (2017) present evidence supporting the logistic model in the case of installed capacity of generation (in GW) of solar and wind. However, our definition of RES is not the installed capacity of generation but the proportion of renewables over the gross generation. Therefore, a more proper model is the Censored Normal Tobit model (CNT) with left censoring of zero and right censoring of one. We estimate the CNT model, but its performance is lower than the linear model[19].

The third target set by CEP2020 is a cut by 20% in gross primary energy consumption[20] by 2020 and by 27% by 2030 compared to projections published in 2007 and in 2009 (EC, 2007, 2009). Projections (baseline case, see Figure 56 on page 58 of EC, 2007) assume an annual increase of electricity demand of 1.5% in the period 2010-2020 and an increase of 0.8% in the period 2020-2030. We use consumption as the measure of electricity demand. We model yearly consumption by using models (5) and (6)[21]. The trend model fits data better, and we generate 100,000 trajectories and compute 1% quantiles of the empirical distribution of terminal values for horizons 2020 and 2030. With 2020, we compare the quantile against the total amount used in 2010 multiplied by $1.015^{10}$. With 2030, we multiply the total amount used in 2010 by $(1.015^{10} * 1.008^{10})$. The target is the reduction of these amounts by 20% (2020) and 27% (2030). Simulation results are summarized in Table 7.[22]

---

[18] As a robustness check we have also computed the probabilities of compliance for the 2020 and 2030 targets using the AR(2) models. Results do not change.
[19] Panel C in Table A4 in the Appendix contains a comparison of the Akaike Information Criteria (AIC) of the trend model and the Censored Normal Tobit model.
[20] We use aggregate data of consumption obtained from ENTSO-E. They include both industry and household consumption. ENTSO-E calculates the values of the national electrical consumption (representing 100% of national values) as follows: National electrical consumption = Net Generation + imports – exports – consumption of pumps.
[21] Estimation results are not included to save space and are available on Appendix, Table A5.

[INSERT TABLE 7 HERE]

Results suggest the probability of achieving CEP's target for horizons 2020 and 2030 are lower than 1% in Austria and Spain.

## 6. Conclusions

The three targets set in EU's CEP2020 package aim to increase the EU's energy security, cut dependence on imported energy, create jobs, advance green growth and make Europe more competitive. The compatibility of these three targets of reducing emissions and consumption and increasing production from renewable sources seems to be taken for granted. This paper studies the extent to which the determinants of wholesale electricity prices and the relationship between wholesale and household retail prices can shed light on this point. The part of generation from renewables associates with decreases in wholesale prices, and is the only factor presenting relevant impact in the ten EU countries studied in the period 2008-2016. Other factors such as fuel prices, weather variables and net balance between exports and imports help in explaining wholesale electricity prices, but the significance of their explanatory power varies across countries.

The impact of renewables on household prices should be negative on the cost of energy part and positive on the government wedge part. We document a positive and relevant effect of renewables on the non-energy component of household electricity prices in six out of ten countries. Besides, in Finland, Portugal, and Spain the government's fiscal position is a leading indicator of the non-energy component of household prices. Increases in public deficits are associated with posterior increases in the taxes and levies included in the government wedge part of household electricity prices. This result remarks the role of policy choices on the observed disconnection in the EU between wholesale prices and final household prices. As decreases in wholesale prices should promote consumption, an apparent contradiction emerges between the target of an increase in renewables and the target of a reduction in consumption. However, the impact of renewables on the non-energy part of household wholesale electricity prices is positive and significant in most countries. Therefore, decreases in wholesale prices, that may compromise the CEP2020 target of the decrease in consumption, do not necessarily translate into lower household retail prices. In competitive markets, reductions in wholesale prices should pass through quickly to retail markets. But policy choices (e.g. support to renewables) may impede this pass-through.

---

[22] Using the estimated AR(2) models we get the results in Table A8 in Appendix A. Results do not change.

Third, simulations-based forecasts suggest that Austria, Portugal, and Spain will struggle in achieving EU's CEP target of reductions in greenhouse gas emissions by 2030 and 2030. With 2030, Austria, France, Germany, Portugal and Spain present probabilities lower than 1% of achieving the cut in GHG emissions consistent with the CEP2020 target. Austria and Spain are unlikely to reduce consumption in the amounts required by 2020 and 2030. Finland and France are unlikely to meet the national targets of production from RES in 2030 or 2030.

Regarding policy recommendations, the current CEP2020 framework is probably too complex and expensive. If the EU climate policy focuses on the abatement of GHG emissions, they should use a single policy instrument (e.g. a carbon tax). The key target should be decreasing carbon consumption and not carbon production. Taxing carbon production will likely incentive carbon-intensive industries to move to more lenient jurisdictions.

This analysis of the impact of renewables on wholesale prices does not separate the effect of different technologies (wind, solar, biomass). As the structure of generation assets varies across countries, this study could be extended by considering a more granular analysis of the impact of each technology separately. On the household prices, in this paper, we consider average-size household consumers. However, there are five types of households, and so we can extend the research to test for differences among consumer types. Analyzing industrial prices is an additional avenue for further research. Improving the simulations is another area of interest, through a more comprehensive integrated model including GHG emissions, consumption and renewables share and the impact of alternative policy choices. This is an immediate extension we left for future research

**Acknowledgments**

We are grateful to Hipòlit Torrò, Julien Chevallier, and participants in 9th Valencia Research Workshop on Energy Markets and CEM2018 Roma Conference for helpful comments and suggestions. Three anonymous referees provided many astute suggestions that considerably improved the manuscript. The usual disclaimer applies. We acknowledge financial support from DGICYT, through grant ECO2016-77807-P and from FUNCAS, through grant PRELEC2020-2017/00085/001.

**Table 1**

Determinants of Wholesale Monthly Electricity Prices

|  | AU | DK | FI | FR | GE | GR | IT | PT | SE | SP | WTp |
|---|---|---|---|---|---|---|---|---|---|---|---|
| **RES** | **-16.72** | **-31.60** | **-42.02** | **-42.79** | **-29.50** | **-16.40** | **-17.51** | **-11.84** | **-26.58** | **-50.61** | 0.000 |
| **BRENTOIL** | **0.112** | -0.114 | -0.166 | **0.260** | 0.042 | **0.196** | **0.298** | 0.024 | -0.168 | 0.073 | 0.000 |
| **COAL** | **0.126** | **0.206** | 0.167 | 0.134 | **0.189** | 0.108 | 0.074 | 0.011 | **0.306** | -0.043 | 0.003 |
| **GAS** | -0.057 | 1.763 | **4.279** | -4.355 | -1.500 | **-4.679** | -3.407 | -2.402 | **4.983** | -2.341 | 0.001 |
| **CARBON** | **0.967** | 0.010 | 0.088 | 0.087 | **0.670** | 0.326 | 0.170 | **0.535** | -0.135 | 0.182 | 0.018 |
| **CONSM** | **0.008** | 0.000 | **0.007** | 0.000 | 0.000 | 0.002 | **0.001** | -0.004 | 0.002 | 0.000 | 0.000 |
| **TEMP** | -0.112 | **-0.934** | **-0.868** | 0.024 | **-0.372** | 0.204 | 0.354 | 0.243 | **-1.102** | 0.352 | 0.006 |
| **NETBAL_DIV** | 0.000 | **0.006** | **0.005** | -0.001 | 0.000 | 0.000 | **0.002** | **-0.006** | -0.004 | 0.002 | 0.000 |
| **DUM_CWE** | 1.311 | 4.027 | 5.175 | -5.724 | 0.401 | -5.338 | -7.882 | 4.563 | 3.428 | 5.143 |  |
| **DUM_NWE** | -2.793 | -3.583 | -4.908 | -5.570 | -1.363 | -6.418 | -6.371 | 2.207 | -3.743 | 2.599 |  |
| **AR(1)** | **0.271** |  |  |  |  |  |  |  |  |  |  |
| **AR(12)** | **-0.235** |  |  |  |  |  |  |  |  |  |  |
| **AR(24)** | **-0.173** |  |  |  |  |  |  |  |  |  |  |
| **Fixed S Effects** | YES | YES | YES | YES | YES | YES | YES | YES | YES | YES |  |
| **Adjusted R2** | 0.894 |  |  |  |  |  |  |  |  |  |  |
| **DW Stat** | 1.961 |  |  |  |  |  |  |  |  |  |  |
| **Sample size** | 1080 |  |  |  |  |  |  |  |  |  |  |
| **Variable** | RES | BRENT | COAL | GAS | CO2 | CONSM | TEMP | NETBAL |  |  |  |
| **VIF** | 5.450 | 2.985 | 7.375 | 3.888 | 5.587 | 4.235 | 6.081 | 2.542 |  |  |  |
| **Hausman Test** | 80.270 |  |  |  |  |  |  |  |  |  |  |

This table presents the results of $p_{t,j} = \beta_{0,j}RES_{t,j} + \sum_{i=1}^{K}\beta_{j,i}X_{t,j,i} + \eta_{t,j}$ where the dependent variable $p_{j,t}$ is the wholesale electricity price in country $j$ at time $t$, and $X_{t,j,i}$ are $i=1,\ldots,K$ explanatory variables either stochastic (e.g. consumption) or deterministic (e.g. dummy seasonal variables). The noise process $\eta_{t,j}$ follows $\phi(B)\eta_{t,j} = \theta(B)\sigma_{t,j}\varepsilon_{t,j}$ $\phi(B) = 1 - \sum_{l=1}^{\phi}\phi_l B^l$; $\theta(B) = 1 - \sum_{l=1}^{\theta}\theta_l B^l$. Functions $\phi(B)$ and $\theta(B)$ are polynomial functions of the backshift operator B, $\sigma_{t,j}$ is the volatility and the innovation $\varepsilon_{t,j}$ is N(0, $\sigma_{t,j}$). Parameters in **boldface** are significant at 1%. Sample is from 1/2008 to 12/2016. Sample size is 1,080 observations. WTp is the p-value of the Wald test.

**Table 2**

Household Retail Prices (Government Wedge)

|  | AU | DK | FI | FR | GE | GR | IT | PT | SE | SP | WTp |
|---|---|---|---|---|---|---|---|---|---|---|---|
| **RES** | -10.679 | **83.382** | **119.713** | 106.208 | 50.421 | 48.423 | **199.856** | **41.802** | 55.473 | **22.253** | 0.0000 |
| **D(PB(-1))** | -0.823 | 0.483 | **-1.703** | -0.494 | -0.764 | 0.124 | -2.949 | **-2.421** | 0.877 | **-1.145** | 0.0000 |
| **AR(1)** | **0.853** | | | | | | | | | | |
| **Adjusted R2** | 0.997 | | | | | | | | | | |
| **DW Stat** | 2.048 | | | | | | | | | | |
| **Sample size** | 180 | | | | | | | | | | |
| **Variable** | RES | D(PB) | | | | | | | | | |
| **VIF** | 1.174 | 1.140 | | | | | | | | | |
| **Hausman Test** | 70.260 | | | | | | | | | | |

This table presents the results of (4) $(h_{t,j} - p_{t,j}) = \alpha_j + \delta_j RES_{t,j} + \varphi_j \nabla PB_{t-1,j} + \eta_{t,j}$. The dependent variable $(h_{j,t} - p_{t,j})$ is the household retail electricity price minus the wholesale price in country $j$ at time $t$, and the noise process $\eta_{t,j}$ follows (2)-(3). RES is the generation by renewables divided by the gross generation and $\nabla PB$ is the change in government deficit-to-GDP ratio (positive if government budget surplus). As in (1), we test whether the slope corresponding to the explanatory variable is the same for all countries by means of a Wald test. Parameters in **boldface** are significant at 1%. Sample is biannual data from 2008 to 2016. Sample size is 180 observations.

**Table 3**

GHG Emissions in 2020 and 2030

|  | AU | DK | FI | FR | GE | GR | IT | PT | SE | SP |
|---|---|---|---|---|---|---|---|---|---|---|
| **GHG emissions 1990** | 79700 | 72104 | 72307 | 555771 | 1262988 | 105577 | 524115 | 61133 | 729905 | 293449 |
| **Target 2020** | 63760 | 57683 | 57846 | 444617 | 1010391 | 84462 | 419292 | 48906 | 583924 | 234760 |
| **Q(1%,2020)** | **70379** | 37049 | 43005 | 415216 | 854874 | 68813 | 331350 | **56697** | 455318 | **266712** |
| **Q(50%,2020)** | 75341 | 41855 | 50723 | 440242 | 900141 | 72472 | 370286 | 62955 | 50677154 | 299995 |
| **Target 2030** | 47820 | 43263 | 43384 | 333463 | 757793 | 63346 | 314469 | 36680 | 437943 | 176070 |
| **Q(1%,2030)** | **60744** | 15604 | 23329 | **337625** | **793476** | 18555 | 192723 | **45986** | 335635 | **185767** |
| **Q(50%,2030)** | 65683 | 20388 | 31010 | 362531 | 838516 | 22196 | 231472 | 52214 | 38684016 | 218890 |

The table shows the results of an analysis of 100,000 trajectories. We model yearly GHG emissions in all countries by using a linear trend model $GHG_t = \beta_0 + \beta_1 t + \varepsilon_t$. Based on this model we compute 1% quantiles of the empirical distribution of terminal values for horizons 2020 and 2030. **Boldface** figures mean that the country has less than 1% probability of meeting the target. Units are Thousand Tones.

**Table 4**
Average Growth Rates 2008-2016

|  | AU | DK | FI | FR | GE | GR | IT | PT | SP | SE | Total |
|---|---|---|---|---|---|---|---|---|---|---|---|
| **GHG** | -1.3% | -3.0% | -2.2% | -1.3% | -1.0% | -4.6% | -2.7% | -1.7% | -2.4% | -1.6% | -2.2% |
| **CONSM** | 0.9% | -1.1% | -0.3% | -0.3% | 0.3% | -1.2% | -1.2% | -0.3% | 0.1% | -1.5% | -0.5% |
| **RES** | 1.8% | 10.9% | 2.5% | 3.5% | 6.9% | 15.8% | 7.1% | 7.9% | 10.0% | 1.0% | 6.7% |

The table shows the average yearly growth rates for GHG emission, CONSM (consumption) and RES (renewable energy sources). The time span is from 01/2008 to 12/2016.

**Table 5**
GHG Emissions in 2020 and 2030. Model (7).

| | | AU | DK | FI | FR | GE | GR | IT | PT | SP | SE |
|---|---|---|---|---|---|---|---|---|---|---|---|
| | **GHG in 1990** | 79700 | 72104 | 72307 | 555771 | 1262988 | 105577 | 524115 | 61133 | 293449 | 72991 |
| | **GHG in 2008** | 89128 | 68453 | 72975 | 541886 | 1000000 | 134630 | 557990 | 79335 | 421075 | 65348 |
| | **TARGET 2020** | 63760 | 57683 | 57846 | 444617 | 1010391 | 84462 | 419292 | 48906 | 234760 | 58392 |
| | **TARGET 2030** | 47820 | 43263 | 43384 | 333463 | 757793 | 63346 | 314469 | 36680 | 176070 | 43794 |
| Panel A | Q(1%, 2020) | **68463** | 45672 | 52068 | 415180 | 784830 | 79240 | 382330 | **58972** | 295760 | 48908 |
| Panel A | Q(50%, 2020) | 75905 | 50637 | 57728 | 460310 | 870140 | 87853 | 423890 | 65382 | 327910 | 54224 |
| Panel A | Q(1% 2030) | 53959 | 35996 | 41037 | 327220 | 618560 | 62452 | 301330 | **46478** | **233100** | 38546 |
| Panel A | Q(50% 2030) | 65468 | 43674 | 49790 | 397020 | 750490 | 75773 | 365600 | 56392 | 282820 | 46768 |
| Panel B | Q(1%, 2020) | 62767 | 41872 | 47735 | 380640 | 719530 | 72646 | 350520 | **54065** | 271150 | 44838 |
| Panel B | Q(50%, 2020) | 69359 | 46270 | 52749 | 420610 | 795090 | 80276 | 387330 | 59743 | 299630 | 49548 |
| Panel B | Q(1%, 2030) | 39524 | 26366 | 30059 | 239680 | 453080 | 45745 | 220720 | 34044 | 170740 | 28234 |
| Panel B | Q(50%, 2030) | 47608 | 31759 | 36207 | 288710 | 545750 | 55101 | 265860 | 41007 | 205660 | 34009 |
| Panel C | Q(1%, 2020) | 59561 | 39734 | 45298 | 361200 | 682780 | 68937 | 332620 | **51304** | **257300** | 42549 |
| Panel C | Q(50%, 2020) | 66150 | 44129 | 50309 | 401160 | 758310 | 76563 | 369410 | 56979 | 285770 | 47255 |
| Panel C | Q(1%, 2030) | 33332 | 22236 | 25350 | 202140 | 382110 | 38579 | 186140 | 28711 | 144000 | 23812 |
| Panel C | Q(50%, 2030) | 40425 | 26968 | 30744 | 245150 | 463410 | 46788 | 225750 | 34821 | 174640 | 28878 |

The table shows the results of the analysis of 100,000 simulated trajectories. We model yearly GHG emissions in all countries by the pool regression $GHG_{i,t} = \beta_{i,0} + \beta_{i,1} CONS_{i,t} + \beta_{i,2} RES_{i,t} + \varepsilon_{i,t}$. Based on this model we compute 1% quantiles of the empirical distribution of the terminal values for horizons 2020 and 2030. Panel A assumes that both the consumption variable and the proportion of renewables will continue to evolve at the average variation rate in 2008 and 2016. Panel B presents a scenario in which consumption decreases by 2% per year and RES increases by 2% yearly. Panel C show a scenario of which consumption decreases by 2% per year and RES increases by 6% yearly. Boldface figures mean that the targets have less than 1% chance of being achieved. Units are Thousand Tones.

**Table 6**

National overall targets for the share of energy from renewable sources in gross final consumption of energy in 2020 and 2030.

|  | AU | DK | FI | FR | GE | GR | IT | PT | SE | SP |
|---|---|---|---|---|---|---|---|---|---|---|
| **Target 2020** | 34% | 30% | 38% | 23% | 18% | 18% | 17% | 31% | 49% | 20% |
| Q(1%,2020) | 49% | 32% | 11% | 11% | 19% | 24% | 21% | 34% | 41% | 25% |
| Q(50%,2020) | 69% | 56% | 24% | 17% | 32% | 36% | 34% | 48% | 50% | 36% |
| **Q(99%,2020)** | 89% | 81% | **37%** | **23%** | 46% | 47% | 47% | 61% | 59% | 48% |
| **Target 2030** | 34% | 30% | 38% | 27% | 27% | 27% | 27% | 31% | 49% | 27% |
| Q(1%,2030) | 48% | 36% | 11% | 10% | 32% | 27% | 21% | 33% | 40% | 24% |
| Q(50%,2030) | 69% | 64% | 24% | 17% | 64% | 40% | 34% | 48% | 53% | 36% |
| **Q(99%,2030)** | 89% | 92% | **37%** | **23%** | 98% | 53% | 48% | 62% | 66% | 48% |

The table shows the results of an analysis of 100,000 simulated trajectories. We model yearly production by renewables in all countries by using linear trend models. Based on this model we compute 1%, 50%, and 99% quantiles of the empirical distribution of RES terminal values for horizons 2020 and 2030 and compare the 99% quantile against CEP targets.

**Table 7**

Projections and Targets of Electricity consumption in 2020 and 2030

|  | **AU** | **DK** | **FI** | **FR** | **GE** | **GR** | **IT** | **PT** | **SE** | **SP** |
|---|---|---|---|---|---|---|---|---|---|---|
| **Gross Inland Consumption 2010** | 68324 | 841521 | 87467 | 513292 | 547422 | 53551 | 330455 | 52206 | 147090 | 260609 |
| **Projection 2020** | 79293 | 976620 | 101509 | 595696 | 635306 | 62148 | 383507 | 60587 | 170704 | 302447 |
| **Target 2020** | 63434 | 781296 | 81207 | 476557 | 508244 | 49718 | 306805 | 48470 | 136563 | 241958 |
| **Q(1%,2020)** | **70820** | 686892 | 78087 | 438167 | 501012 | 43970 | 279516 | 45724 | 116115 | **255171** |
| **Q(50%,2020)** | 73779 | 720671 | 82612 | 465609 | 537639 | 46968 | 294915 | 47673 | 125437 | 265633 |
| **Projection 2030** | 85870 | 1057623 | 109928 | 645105 | 687999 | 67303 | 415315 | 65612 | 184863 | 327533 |
| **Target 2030** | 62685 | 772065 | 80248 | 470926 | 502239 | 49131 | 303180 | 47897 | 134950 | 239099 |
| **Q(1%,2030)** | **76660** | 586028 | 75772 | 411723 | 497497 | 37471 | 245415 | 43040 | 94575 | **261294** |
| **Q(50%,2030)** | 79604 | 619645 | 80274 | 439033 | 533948 | 40454 | 260739 | 44980 | 103852 | 271705 |

The table shows the results of an analysis of 100,000 simulated trajectories. We model yearly consumption in all countries by using linear trend models. We compute 1% quantiles of the empirical distribution of terminal values for horizons 2020 and 2030. In the cases of 2020 and 2030, we compare these quantiles against Projections 2020 and 2030, computed using gross inland consumption used in 2010 multiplied by $1.015^{10}$ and by $(1.015^{10} * 1.0081^0)$ respectively. We then compare against Targets 2020 and 2030 which are reductions of this projected amount by 20% and 27% respectively. Units are GWh.

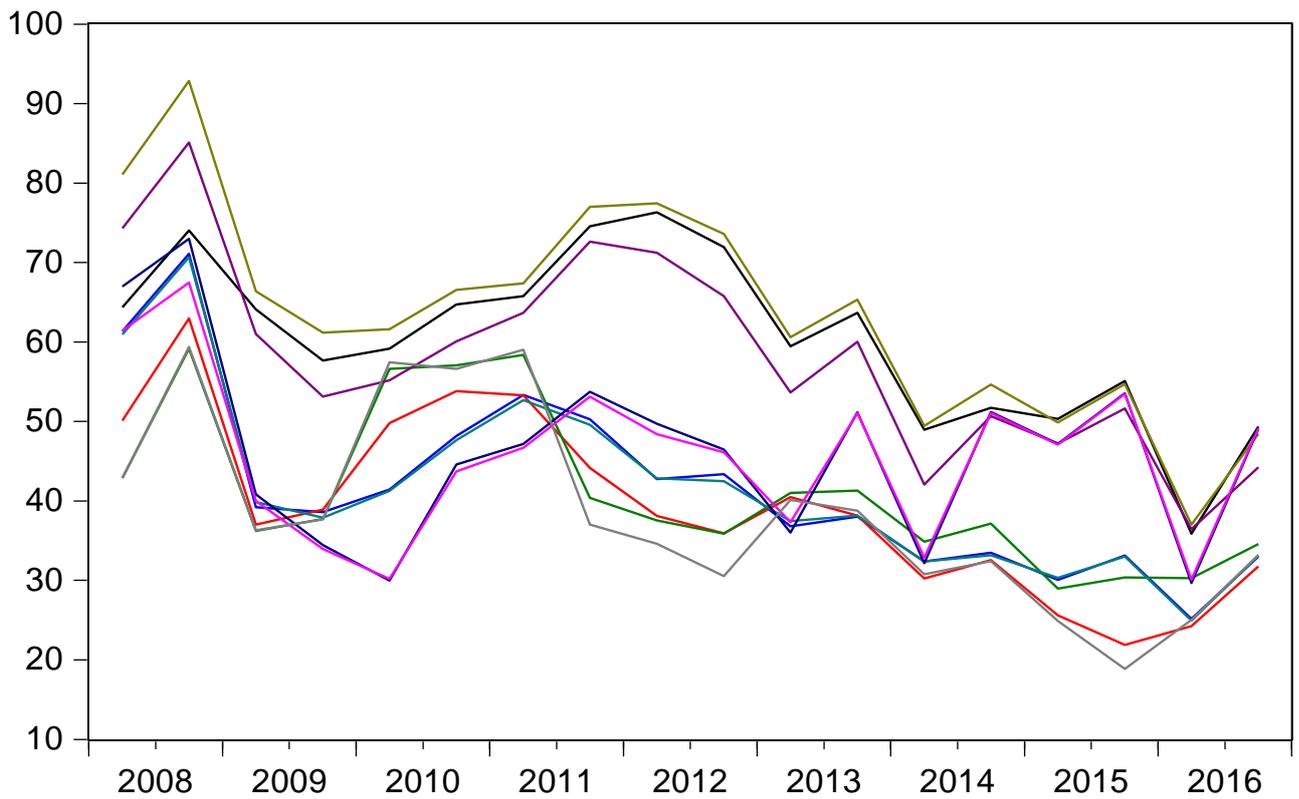

**Figure 1: Biannual Wholesale Electricity Prices**

**Figure 2: Biannual Household Retail Electricity Prices**

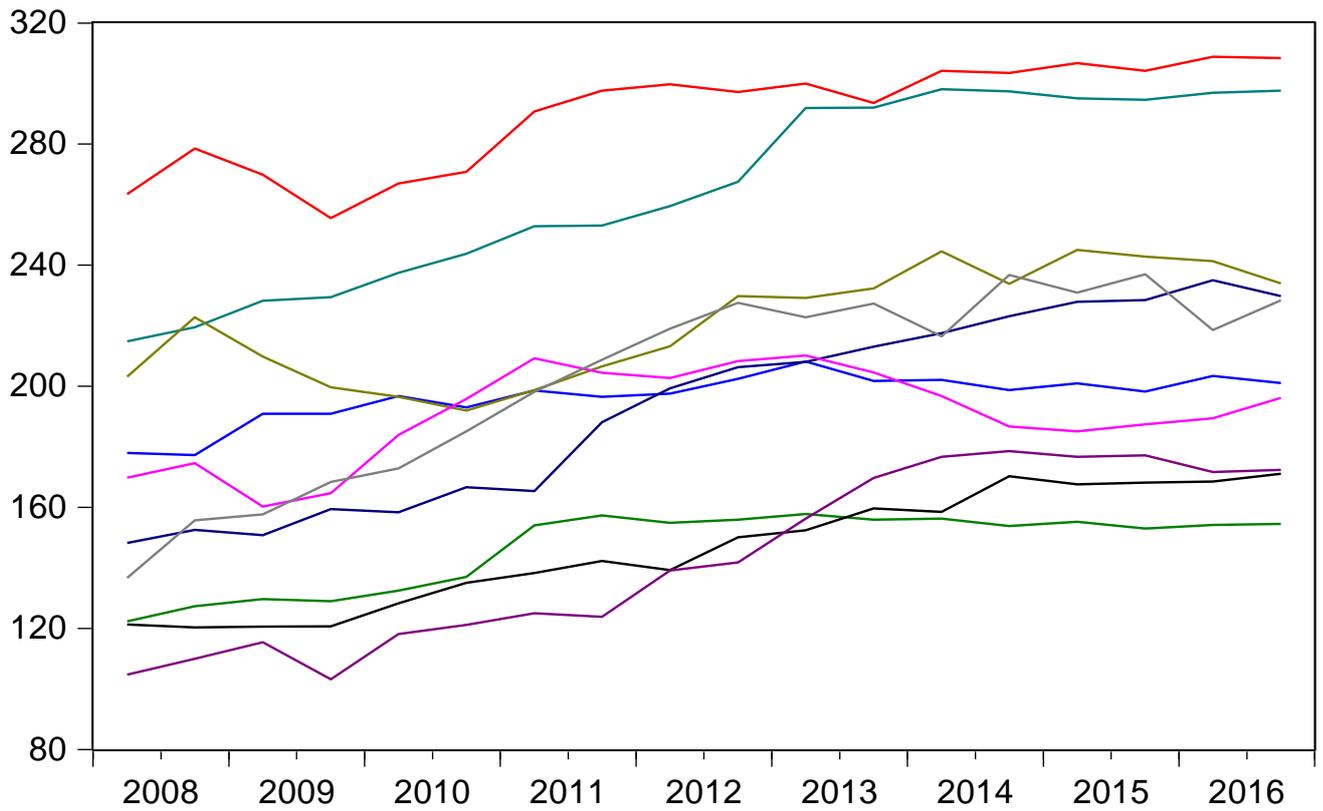